\documentclass[acmlarge]{acmart}

\usepackage{csquotes} 
\usepackage{booktabs}
\usepackage{graphicx}

\AtBeginDocument{%
  \providecommand\BibTeX{{%
    \normalfont B\kern-0.5em{\scshape i\kern-0.25em b}\kern-0.8em\TeX}}}

\setcopyright{acmlicensed}
\acmJournal{JOCCH}
\acmYear{2023} \acmVolume{1} \acmNumber{1} \acmArticle{1} \acmMonth{1} \acmPrice{15.00}\acmDOI{10.1145/3627167}

\acmJournal{JOCCH}

\usepackage{color}
\definecolor{RED}{rgb}{0.9,0,0}
\definecolor{GREEN}{rgb}{0,0.8,0}
\definecolor{BLUE}{rgb}{0,0,0.8}
\definecolor{MAGENTA}{rgb}{0,0.7,0.7}
\definecolor{PURPLE}{rgb}{0.4, 0.0, 0.6}
\definecolor{GRAY}{rgb}{0.5, 0.5, 0.5}

\begin{document}

\title{Write What You Want: Applying Text-to-video Retrieval to Audiovisual Archives}

\author{Yuchen Yang}
\email{yuchen.yang@epfl.ch}
\affiliation{%
  \institution{École Polytechnique Fédérale de Lausanne - Laboratory for Experimental Museology+}
  \streetaddress{Rue des Jordils 41}
  \city{St-Sulpice VD}
  \state{Lausanne}
  \country{Switzerland}
  \postcode{1025}
}

\renewcommand{\shortauthors}{Yang}

\begin{abstract}
Audiovisual (AV) archives, as an essential reservoir of our cultural assets, are suffering from the issue of accessibility. The
complex nature of the medium itself made processing and interaction an open challenge still in the field of computer vision,
multimodal learning, and human-computer interaction, as well as in culture and heritage. In recent years, with the raising of
video retrieval tasks, methods in retrieving video content with natural language (text-to-video retrieval) gained quite some
attention and have reached a performance level where real-world application is on the horizon. Appealing as it may
sound, such methods focus on retrieving videos using plain visual-focused descriptions of what has happened in the video
and finding videos such as instructions. It is too early to say such methods would be the new paradigms for accessing and
encoding complex video content into high-dimensional data, but they are indeed innovative attempts and foundations to build
future exploratory interfaces for AV archives (e.g. allow users to write stories and retrieve related snippets in the archive, or
encoding video content at high-level for visualisation). This work filled the application gap by examining such text-to-video
retrieval methods from an implementation point of view and proposed and verified a classifier-enhanced workflow to allow
better results when dealing with in-situ queries that might have been different from the training dataset. Such a workflow is
then applied to the real-world archive from Télévision Suisse Romande (RTS) to create a demo. At last, a human-centred
evaluation is conducted to understand whether the text-to-video retrieval methods improve the overall experience of accessing
AV archives.

\end{abstract}




\maketitle

\section{Introduction}
Audiovisual (AV) archives comprise a large and vital part of our cultural legacy. However, the increasing volume and the lack of suitable interfaces and methods for accessing content pose immense challenges for AV archives to realise their public value \cite{edmondson2004audiovisual}. To solve this issue, efforts from different perspectives are made. On the content end, archivists have discussed in detail the interrelationships between the analogue and the digitised versions of AV materials, as well as how to form digitisation strategies carefully \cite{cavallotti2018grain} to enhance preservation and accessibility in the digital age. As a growing number of new technologies (in computer vision and natural language processing) becomes available, some scholars and practitioners start to advocate that cultural content is cultural data \cite{manovich2020cultural, padilla2018collections}, and that understanding and display of cultural phenomena could be done in a data-driven way and at a larger scale. 

On top of the complexity of the content, copyright, bias brought by archiving and preservation (western and English focused), interface and technology (not everyone knows how to use a web interface or have internet access), and curatorial decision (focus on the popular object and theme but not the long tail) also contribute to the limited access. To improve the accessibility from the consumer end, some memory institutions have started experimenting with computationally-enhanced methods to improve the traditional search (fuzzy search, content recommendation, etc.) or build novel explorative interactions (immersive, embodied, interactive visualisations) \cite{kenderdine2021experimental}. Although not fully adapted nor matured, these newly emerged efforts actively probe into the unknown territory beyond keyword search and themed exhibitions. 

The combination of a datafied archive and novel method for access aims to "open up" the full potential of memory materials to the general public and to transform the access to archives to be more democratic, personalised, and explorative \cite{kenderdine2021computational}. T\_Visionarium\footnote{\url{https://www.jeffreyshawcompendium.com/portfolio/t_visionarium-ii/}}, as one of the earliest experiments in this direction, uses properties such as emotion, action, and gender (annotated by hand) to connect the content and allows the audience to explore and build clip-chains based on these properties. In contrast, the more recent SEMIA project \cite{masson2020exploring} works on experimenting with alternative AV archive interfaces for exploring through features such as colour, shape, and visual clutter and encourages the serendipitous discovery of an archive rather than search. At the core of these attempts to improve accessibility for cultural and heritage archives is innovative and appropriate methods for encoding content. How to apply state-of-the-art computational methods to extract less explored features in AV archives (such as gestures, natural language descriptions, and other abstract-level semantic embeddings) to support new applications becomes a pivotal question to answer. 

Text-to-video retrieval models attack the multimodality of AV content and aim to facilitate the video search most intuitively - using arbitrary natural language sentences. Unlike unimodal tasks (e.g. image retrieval, focus only on visual modality), text-to-video retrieval is such a challenging task since it not only requires a solid understanding of the text query and the multimodal content of videos but, more importantly, their inter-modal correlation \cite{yang2021taco}. The ultimate goal of such methods is to allow people to query and retrieve video clips based on any arbitrary narrative sentence. The freedom to use someone's own words to explore similar happenings in a historical archive enables users to explore without prior knowledge and discover serendipitous historical and cultural facts. It would tremendously help create their understanding and narrative towards content, cultivating a more mesmerising sense-making experience \cite{thompson2017logic}.

However, the gap between the state-of-the-art model and the real-world application is not bridged. Looking at proposed datasets for text-to-video retrieval tasks, the text-video pairs are vastly focused on plain visual descriptions. Like any machine learning task, such simplicity is necessary for annotation and model development to be feasible. And it is not deniable that such kind of description works well for certain types of videos (such as how-to or instruction videos). Nonetheless, these models might not suffice the use case speculated where users could freely explore similar happenings in a historical archive using their own words, where user-generated narrative writings are guaranteed to be more complex. 

To better understand the nature of narrative writing and the complexity of queries to expect on AV archives (such as broadcasting archives recording everyday life), this work resorts to the simplest form of such - diaries \cite{lejeune2009diary}. They are mostly simple, straightforward, and detailed in descriptive narration for daily happenings, where location, time, people, quotes, feelings, and emotions are often the key elements \cite{fullwood2009blog} (These traits are verified by randomly selecting 50 diaries and journal entries from the historical diaries and journal archives: Aunt Hattie's Diary\footnote{\url{http://www.aunthattiesdiary.com/}} and Memoirs \& Diaries\footnote{\url{http://firstworldwar.com/diaries/index.htm}}). Amongst all these elements, quotes are the most neglected in the standard datasets available for text-to-video retrieval tasks. The absence of quotes in the textual descriptions makes the model fundamentally weak in handling queries with quotes. However, since speeches in videos do not always correspond to the visual elements, adding quotes to the video-text pair may add noise to the whole process and eventually undermine the performance.
\begin{figure}
    \centering 
    \includegraphics[width=.8\columnwidth]{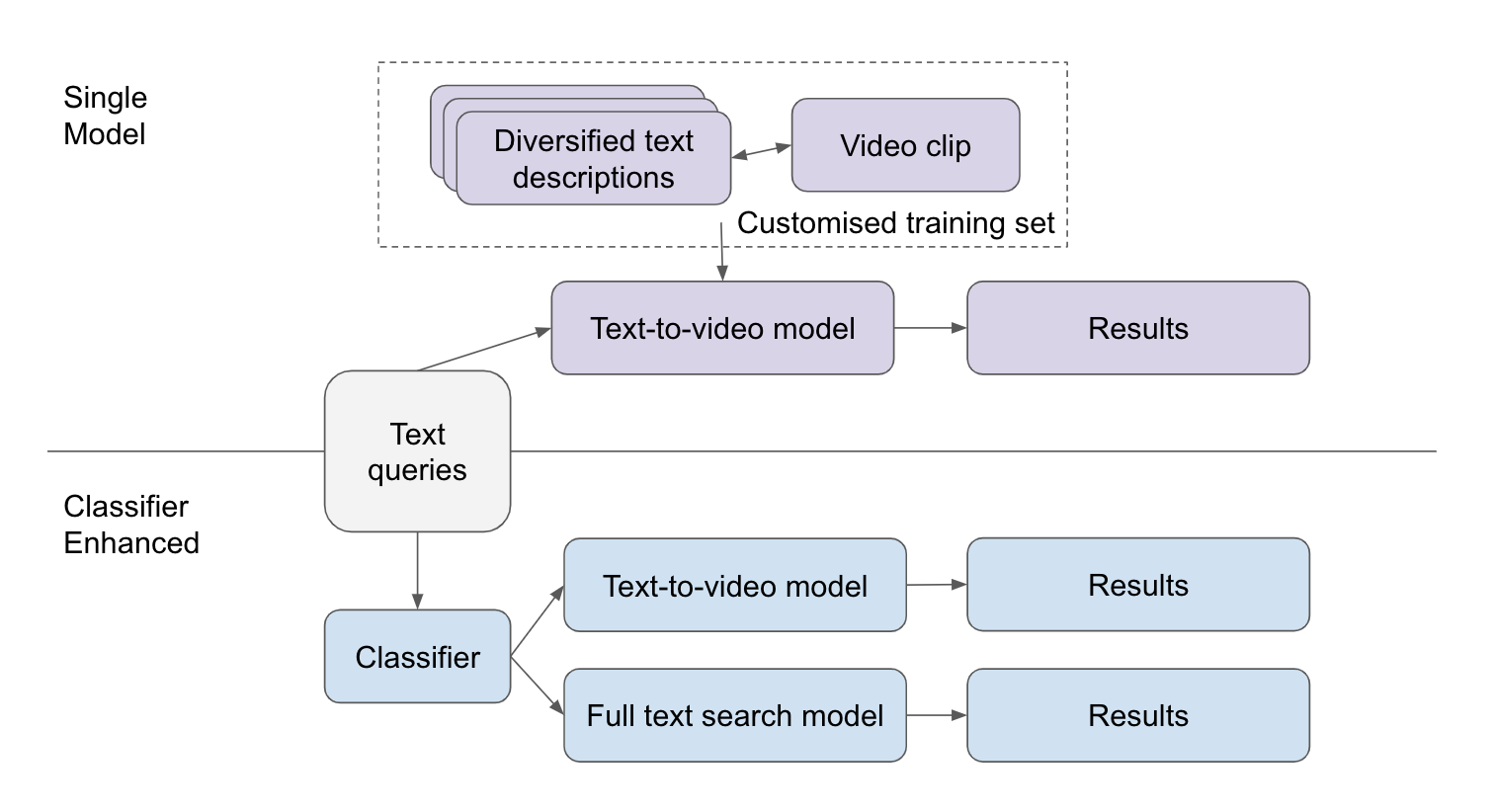} 
    \caption{Overview of the two proposed methods to handle extra query types.}
\label{2}
\end{figure}
This work examined the rising text-to-video retrieval method with a focus on application. We experimented, compared and validated two potential ways to flexibly handle more complex query situations (adding quote-related queries). Fig.\ref{2} illustrates the proposed two methods at a high level. The single model method uses machine learning models to introduce new information (speech) to the original annotation. The enriched annotations serve as a customised training set to understand whether adding information to the annotation would help the model perform better. The classifier-enhanced method uses a classifier to categorise the different types of user input and then send queries to the relevant retrieval model. This method helps to understand whether pre-processing user input would improve overall performance. These two methods were evaluated using standard retrieval datasets and real-world archival content provided by Télévision Suisse Romande (RTS). This work aims to provide a glimpse into the application and improvement of text-to-video retrieval models for memory institutions and open the door for building innovative and explorative interfaces to access archival content in the future. Section 2 of this paper introduces the related work. Section 3 introduces the two proposed methods, and details on evaluation and implementation. Section 4 reports the results and the comparison of the two methods to the state-of-the-art. Section 5 elaborates on the limitation of the work and suggests several directions to future investigate how to close the gap between the state-of-the-art model and applications.

\section{Related Work}
\label{lbl:related-work}

\subsection{Text-to-video retrieval}
\begin{figure}
    \centering 
    \includegraphics[width=.4\columnwidth]{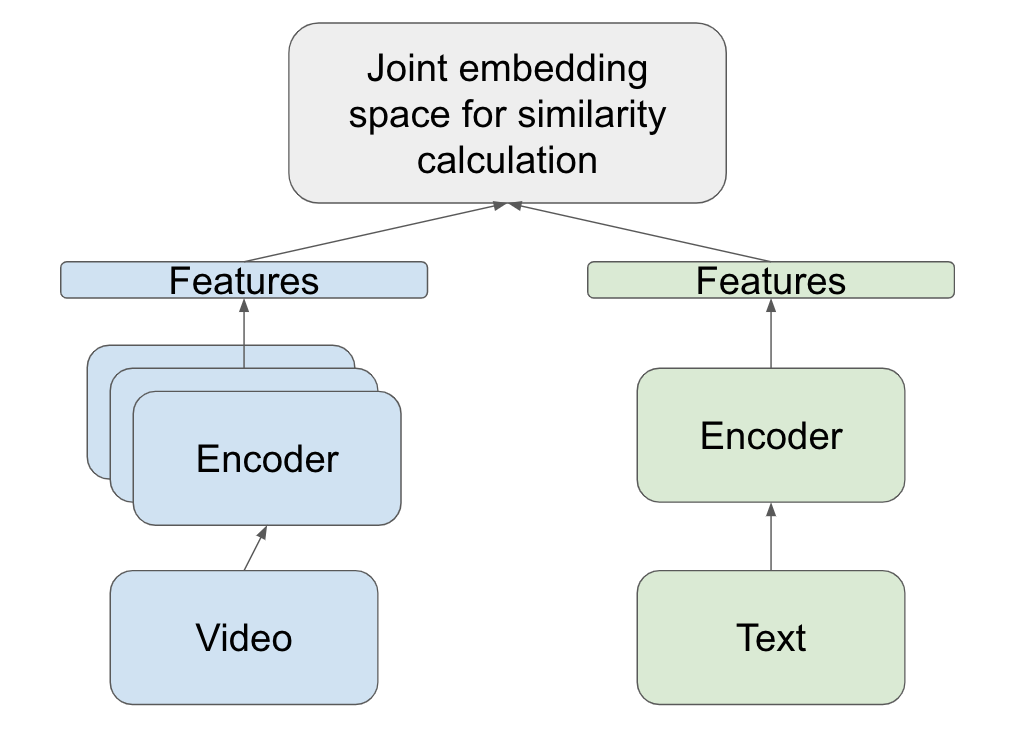} 
    \caption{Higher-level mechanism illustration of classic text-to-video retrieval models.}
\label{1}
\end{figure}

The recently popularised text-to-video retrieval task is at the core of this proposed work. This task takes an arbitrary text query and searches for the most relevant video clips accordingly.  First proposed in 2016 \cite{rohrbach2017movie}, the classic works in text-to-video retrieval tackle the problem with two-stream architecture \cite{miech2019howto100m, miech2020end}. Fig.\ref{1} illustrates the higher-level classic text-to-video retrieval mechanism.

These methods usually utilize videos and their corresponding annotation texts. Videos are transformed by encoders focusing on different video modalities to vectors in the latent space. Text annotations, on the other hand, are also transformed into vectors in the latent space. The features gained from texts and videos are then projected into a common feature space. The paired relationships of the vectors are then used for training (or transforming) this common feature space by the model, so that the paired vectors are always closer to each other. This trained common feature space, called joint embedding space, is then used for future inference and retrieval.

Bi-directional loss \cite{karpathy2014deep}, symmetric cross entropy loss\cite{wang2019symmetric}, and triplet loss \cite{schroff2015facenet} are often used to train a joint embedding space on the assumption that the paired video and text encoding should be closer to each other, hence more similar. The training of text-to-video retrieval models benefits from manually labelled datasets for language-to-video related tasks \cite{rohrbach2017movie, anne2017localizing, xu2016msr, chen-dolan-2011-collecting}. Some works focus on taking advantage of the multimodality of the video and propose a more comprehensive encoding from the video side utilising multimodal cues such as the face, audio, speech \cite{shvetsova2022everything, gabeur2020multi}. 

Aside from research on embedding-based solutions, various methods have been proposed to generalise such models with alternative datasets for pretraining. \cite{
miech2020end} trained on an automated constructed large dataset using transcriptions as text descriptions for video clips. Other works \cite{dzabraev2021mdmmt, portillo2021straightforward, xue2022clip} rely on pre-existing large language to image models to pretrain, and finetune the model using manually labelled text-video datasets. Most recently, \cite{kunitsyn2022mdmmt} proposed a new model that utilises not only the automated generated weakly-paired text-video datasets but also manual-labelled text-image and standard text-video datasets.

While solutions for text-to-video retrieval tasks are diverse and maturing, it is still a bit far from being able to support real arbitrary queries. Most existing state-of-the-art models perform well on simpler text queries, such as plain and visual descriptions. But the performance drops when evaluated on datasets with more sophisticated text descriptions for videos as ground truth \cite{ji2022cret}. A recent study has proven that a better annotation would improve models' performance \cite{chen2022msr}. While most efforts are made to improve encoding or training strategy, little attention is paid to the impact of having an appropriate annotation strategy for a training dataset. 

\subsection{Dialogue act recognizer}
Aside from the text-to-video retrieval method, which targets a one-stop solution to solve the complex natural-language-based retrieval problem, works on chatbot pointed out another way to approach. Inputs to a chatbot system can vary, some seeking specific information, some just for casual chat. To handle the wide variety of input, chatbot systems often implement an intent or act recognition step so that the different needs are properly handled \cite{setyawan2018comparison, schuurmans2019intent}.

The goal of the recognizer is to assign classification labels to each utterance in a conversation. Early works for short text classification use techniques such as bag-of-word (BoW), naive Bayes (NB) and support vector machines (SVM) \cite{wang2012baselines}. Others attempt to improve the performance by leveraging the text with its context \cite{khanpour2016dialogue, bothe2018context}, introducing Conditional Random Field (CRF) to understand the inter-tag dependencies \cite{li2018dual, raheja2019dialogue}.

Such a systematic approach could potentially help to achieve the ultimate goal of using arbitrary text to retrieve video clips in the more complex application settings. Understanding what kind of query the input text is would greatly ease the complexity of the problem. Instead of building an all-in-one model, the solution could be a system with several different retrieval models, and only appropriate ones would be activated when queried. The dialogue act recognizer's goal for this research is to distinguish the type of input query so that it can be sent to the right component.

\section{Methodological Approach}
\label{lbl:methodology}

\subsection{Base text-to-video embedding model}
This work evaluates the difference in performance for the two proposed methods of video retrieval with more complex queries. The minimum requirement for the text-to-video models in use is to consider the speech or audio modality when encoding the video content. In other words, state-of-the-art models, that do not consider audio information at all would not be suitable for such a task. On top of this premise, models used for this research are selected following these principles: representativeness, source code availability, and performance ranking in text-to-video retrieval tasks. In the end, this work chooses the most classical multimodal retrieval model MTT \cite{gabeur2020multi} and the latest state-of-the-art multimodal retrieval model MFT \cite{shvetsova2022everything} as the core methods to construct the two methods for evaluation. MTT utilises Motion, Audio, Scene, OCR, Face, Speech, and Appearance features from the video content, and uses a weighted similarity estimation with bi-directional max-margin ranking loss for training the joint embedding space. MFT focuses on the visual features obtained through ResNet-152 AND ResNeXt-101 backbone and audio features from a trainable CNN with residual layers. For training the joint embedding space, MFT uses a combinatorial loss.

\subsection{Base dataset}
One of the most used standard datasets for video retrieval tasks, MSR-VTT \cite{xu2016msr}, is selected as the basis for this research. This dataset contains videos in various categories, such as music, sports, news, TV shows, movie and drama. The vastly diversified content mirrors real-world heritage AV archives the most. This dataset provides 10,000 video clips harvested from random internet sites, totalling 41.2 hours. Each video clip within this dataset is paired with 20 human annotations, contributing to 200,000 clip-sentence pairs.

\subsection{Single model with enhanced description workflow} 
\subsubsection{Customised training set}
\begin{figure}[h]
    \centering 
    \includegraphics[width=.9\columnwidth]{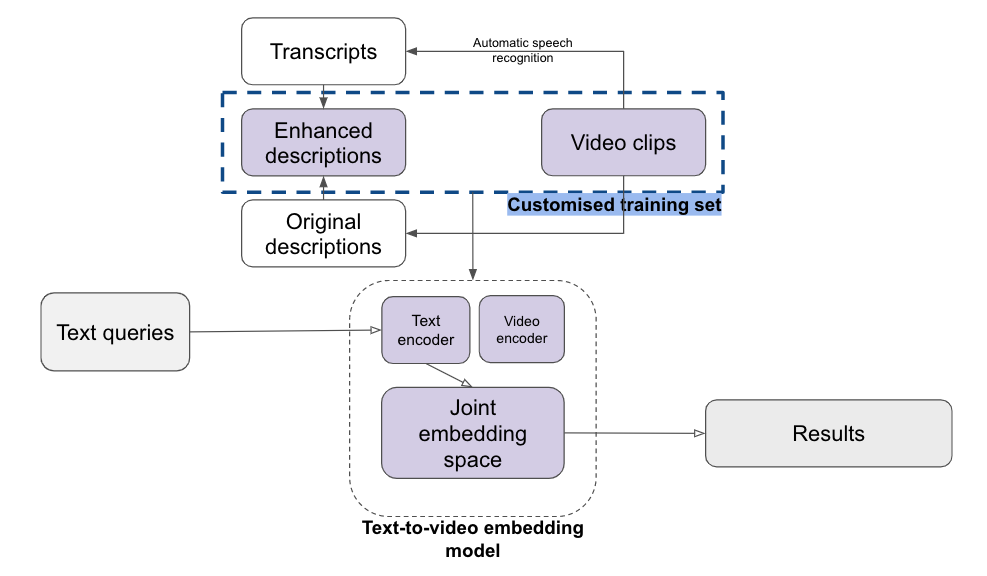} 
    \caption{A detailed look at the proposed single model with enhanced description workflow, the key component - customised training set - of this workflow is highlighted in the blue dashed box.}
\label{7}
\end{figure}

Fig.\ref{7} depicts the single model with enhanced description workflow from end to end. The enhanced descriptions for each video clip are at the core of this workflow. The original annotation in the MSR-VTT dataset for training and testing is limited to plain and visual descriptions. For example, for a video clip of a short conversation between a girl and judges on a singing competition show, the original descriptions, such as "a girl and the judges talking on the voice" and "a girl is talking to the judges on a game show", do not reflect on the content of the conversation at all. To better understand if a more diverse annotation in the dataset would improve the performance for complex queries, this workflow focuses on building a customised training set with the enhanced descriptions following a previous work using automatic speech recognition (ASR) \cite{miech2019howto100m}. The customised training set is contracted by using the obtained transcripts of each video clip to randomly replace one or more of the 20 original descriptions paired with a video clip, depending on availability.

\subsubsection{Workflow implementation details}
\textbf{Transcripts. }The entire MSR-VTT dataset is fed to the recently proposed Whisper \cite{radford2022robust} model to obtain transcripts for each video clip, using the provided "small model" and following the implementation guides on hugging face\footnote{\url{https://huggingface.co/openai/whisper-small}}. \textbf{Customised training set. }We adapted the popular 1k-A split on MSR-VTT produced by \cite{yu2018joint} for constructing the customised training set. The obtained transcripts for all 9,000 training clips assigned by 1k-a are used to replace one or more original annotations randomly in the dataset for model training. \textbf{Model Training. }The customised training set is then used to train the two base text-to-video embedding models, MMT and MFT, following their official implementation configurations respectively\footnote{MFT:\url{https://github.com/ninatu/everything_at_once}; MMT: \url{https://github.com/gabeur/mmt}}. This workflow utilises a customised training set and produces two new methods: customised MMT and customised MFT.

\subsection{Classifier-enhanced workflow}
\subsubsection{Classifier}
\begin{figure}[h]
    \centering 
    \includegraphics[width=.9\columnwidth]{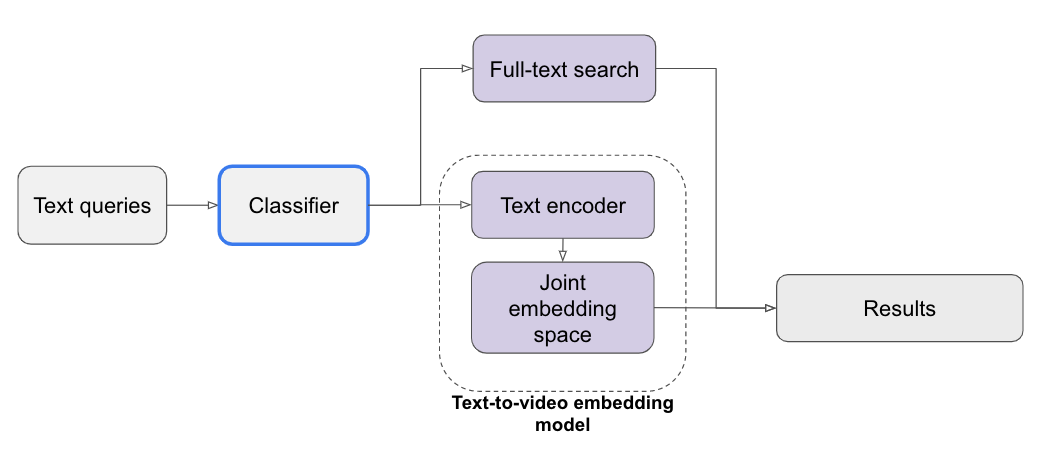} 
    \caption{A detailed look at the proposed classifier-enhanced workflow, the key component - classifier - of this workflow is highlighted in the blue box.}
\label{8}
\end{figure}

As seen in Fig.\ref{8}, the fundamental idea for this workflow is to set a pre-processing step where queries in different types and levels of the complex can be classified and sent to the appropriate retrieval method. Although hard-coded rule sets can function to a certain extent, a machine-learning-based classifier is introduced in the hope of scalability and generalisation. The goal for this classifier at this stage is to distinguish quote or speech-related texts from plain visual descriptions context-free.

Sequence models are utilized in supervised learning scenarios where either the input or output of the model consists of a sequence. Notably, recent advancements in deep learning, particularly in the realm of sequence models, have brought about a significant transformation in the domain of natural language processing (NLP). Prominent examples of sequence models, such as recurrent neural networks (RNNs) and transformers, have consistently performed remarkably on numerous standard NLP benchmarks. We employ a long short-term memory (LSTM) architecture in this workflow to address the classification task.

\subsubsection{Workflow implementation details}
\textbf{Training data for the classifier. }A customised labelled dataset with speeches or quotes and plain visual descriptive sentences is prepared for the training. The labelled data for speeches or quotes are constructed by joining the randomly sampled 2000 sentences from online databases\footnote{\url{https://libguides.bgsu.edu/c.php?g=227160&p=1505718}} using the regular expression filters converted from rules summarised by writing advisors \footnote{\url{https://www.ccis.edu/student-life/advising-tutoring/writing-math-tutoring/introduce-quotations}}, the other 1000 entries in this categories are the random 1000 transcripts from the customised training set in Section 3.3.1. The labelled data for general descriptive sentences are 2000 randomly selected video descriptions from the MSR-VTT 1k-A training set. A random 80-20 split done to produce the training and testing sets. \textbf{Training the classifier. }The binary classifier is trained following previous work\footnote{\url{https://medium.com/holler-developers/intent-detection-using-sequence-models-ddae9cd861ee}} with the labelled training set on the TensorFlow Keras sequential model. It contains an embedding layer representing each word with a vector length of 16. The following 16-unit LSTM layer uses relu activation. The final dense layer has seven units and a softmax activation for classification. The model is fit on the training dataset with a batch size of 32 for seven epochs. \textbf{Text-to-video model. } This workflow sends non-speech or non-quote queries to the text-to-video model. The original MSR-VTT 1k-A training set is used to train the two base text-to-video embedding models MMT and MFT, following their official implementation configurations respectively\footnote{MFT:\url{https://github.com/ninatu/everything_at_once}; MMT: \url{https://github.com/gabeur/mmt}}. \textbf{Full-text search model. }For quote- or speech-related queries, the workflow sends them to the full-text search model, which retrieves videos based on query text and speech-to-text similarity. The speech-to-text data obtained from MSR-VTT in Section 3.3.1 are stored in an ElasticSearch\footnote{\url{https://www.elastic.co/}} database. The similarity is calculated using the ElasticSearch built-in API for full-text queries. This workflow utilises a classifier and produces two new methods based on the two base text-to-video embedding models: classifier MMT and classifier MFT.

\subsection{Comparison with state-of-the-art} 
\subsubsection{Test datasets}
\textbf{Baseline test set. }The popular 1k-A split on MSR-VTT provides the baseline test dataset. \textbf{Customised test set. }A customised test set is introduced to evaluate different methods' performance in complex query situations better. The customised set is constructed on the base of the 1k-A split test set on MSR-VTT. We randomly replaced 50\% of the original ground truth (one random entry from the 20 annotations) with the obtained transcripts for those video clips. The result is 1,000 ground truth pairs with a mixture of plain visual descriptions and speeches or quotes. 
\subsubsection{Items for comparison}
\textbf{Baseline methods. } The original models MMT and MFT were used as baseline methods. \textbf{Enhanced description workflow. }The two new methods - models trained on the customised training set - customised MMT and customised MFT were compared against the baselines. \textbf{Classifier-enhanced workflow. } The two new methods - retrieval workflows with a classifier to pre-process queries based on the two baseline models - classifier MMT and classifier MFT were compared against the baselines.
\subsubsection{Evaluation metrics}
In video retrieval task settings, the performance is normally evaluated by standard metrics "recall at rank N" (R@N where N=1, 5, 10, higher is better) and median rank (MdR, lower is better). However, it is noteworthy that the main focus for such retrieval tasks, in reality, is to explore the archival content rather than finding the exact match. Hence loose indicators like R@5 are likely to be more useful when understanding the performance and hence picked to report in the result.

\subsubsection{Evaluation plan. }
All models were tested without additional pre-training on datasets other than MSR-VTT or the customised MSR-VTT training set to ensure a consistent and comparable result. Table \ref{table:evalplan} provides a more detailed view of the evaluation set-up all for of the six methods mapped in Section 3.5.2. 

\begin{table}[h]
\begin{tabular}{ccc}
\hline
Method & Training Dataset & Testing Dataset \\ \hline
MMT              & Original  & Original + Customised \\
MFT              & Original  & Original + Customised \\\hline
Customised MMT   & Customised & Original + Customised \\
Customised MFT   & Customised & Original + Customised \\ \hline
Classifier MMT   & Original & Original + Customised \\
Classifier MFT   & Original & Original + Customised \\ \hline
\end{tabular}
\caption{Evaluation set-up for comparison with state-of-the-art}
\label{table:evalplan}
\end{table}

\subsection{Human evaluation}
The standard retrieval metric R@N is widely adapted, making it a good indicator for benchmarking different methods. However, One of the main issues is that the testing ground truth is very strict and partial. When calculating for R@N, one ground truth is a pair of one video and one query, and only when that video is retrieved when using that query it counts as a success. This ignores the situation when sometimes the texts or videos are more or less similar, which, in human perception, is considered acceptable results. For example, in ground\_truth\_A, query\_A "a boy crying on the playground" is paired with video\_A (a boy crying on a soccer course). However, in many cases, query\_A may retrieve a video of a teenager crying on a track field. Although the content is extremely similar to the query, it would count as a failure. On the other hand, text-to-video retrieval models have been tested and improved constantly with the standard datasets. Seldom have they been applied to real-world archives. The ability of such models to support a satisfactory experience to explore the archive in the hope of gaining more accessibility and serendipitous findings is untested. We conducted a human evaluation session on the real-world archive RTS to bridge these gaps and provide a more comprehensive view of the methods' performance. For the ease of evaluation, the latest MFT model is used as the base model for text-to-video extraction.

\subsubsection{Dataset. }
Although working with a real-world content partner is a privilege, the reality of RTS being a TV station makes its content organised differently than traditional archives, with very limited content-related metadata. Transforming the archive content to meet the requirements of text-to-video models is necessary. For the ease of the evaluation session, a subset of the whole archive is chosen and pre-processed for fine-tuning. A random batch of 100 videos with a total runtime of around 50 hours is used for comprehensiveness and representativeness. Standard datasets used for Text-to-video retrieval tasks, take MSR-VTT as an example, has an average of 15 seconds in video runtime. For each video, 20 corresponding textual descriptions are attached. The selected videos from RTS are preprocessed to match such statistics.

\subsubsection{Data preprocessing details. }
\textbf{Video Segmentation.} Adaptive Content Detection in PySceneDetect\footnote{\url{https://github.com/Breakthrough/PySceneDetect}} is used for cutting the entire video. Since the result clips vary in duration and can be too short, a 12 seconds threshold is set to group consecutive shot clips into a minimum of 12 seconds. \textbf{Video Captioning.} Although the GPT-powered captioning model (\cite{tewel2022zero}) claims to achieve surprisingly good results, in the scope of archives, descriptions are expected to be accurate. However interesting and useful these models can be for a generative purpose, it is not the most suitable for this task due to the hallucination problem. A dense captioning model, PDVC (\cite{wang2021end}), generates multiple descriptions of the segmented video clips. Although there are random mistakes in captioning, the same level of mistakes is also seen in the MSR-VTT dataset (such as grammar, spelling, and factual mistakes). \textbf{Transcripts. } Similar to Section 3.3.2, videos are sent to the ASR model Whisper to obtain transcripts using their "small model". Since a proportion of the language in the RTS archive is French, these transcripts are then translated into English with Helsinki-NLP/opus-mt-fr-en \footnote{\url{https://huggingface.co/Helsinki-NLP/opus-mt-fr-en}}.

\subsubsection{Single model with enhanced description implementation details. }
\textbf{Customised RTS training set. }We split the preprocess RTS dataset with 80\% for training and 20\% for testing. The obtained transcripts for training clips are used to replace original annotations randomly in the dataset for model training. \textbf{Model Training. }The customised RTS training set is then used to train the two base text-to-video embedding model MFT, following its official implementation configurations. 

\subsubsection{Classifier-enhanced implementation details. }
\textbf{Classifier, full-text search model. }We use the same classifier and full-text search implemented in Section 3.4.2. \textbf{Model training. }The same split for training and testing in producing the customised RTS training set is kept, and the original captions for each video are kept. Following its official implementation configurations, the training split is used to train the two base text-to-video embedding model MFT. This trained MFT will also serve as the baseline method for the evaluation.

\subsection{Evaluation setup}
The objective for the human evaluation part is two-fold. First, it evaluates whether the proposed new methods perform better than the baseline method when handling more diverse queries (quote-related ones, to be specific to this work) from a subjective perspective. Second, it helps to understand whether applying text-to-video retrieval techniques to archives works and if the whole experience counts towards improving the traditional keyword-based search. A total of 11 people (five male and six female, ages between 20 and 35) were recruited through social media to do this evaluation. The evaluation is conducted via Zoom individually. 

\begin{figure}[h]
    \centering 
    \includegraphics[width=.9\columnwidth]{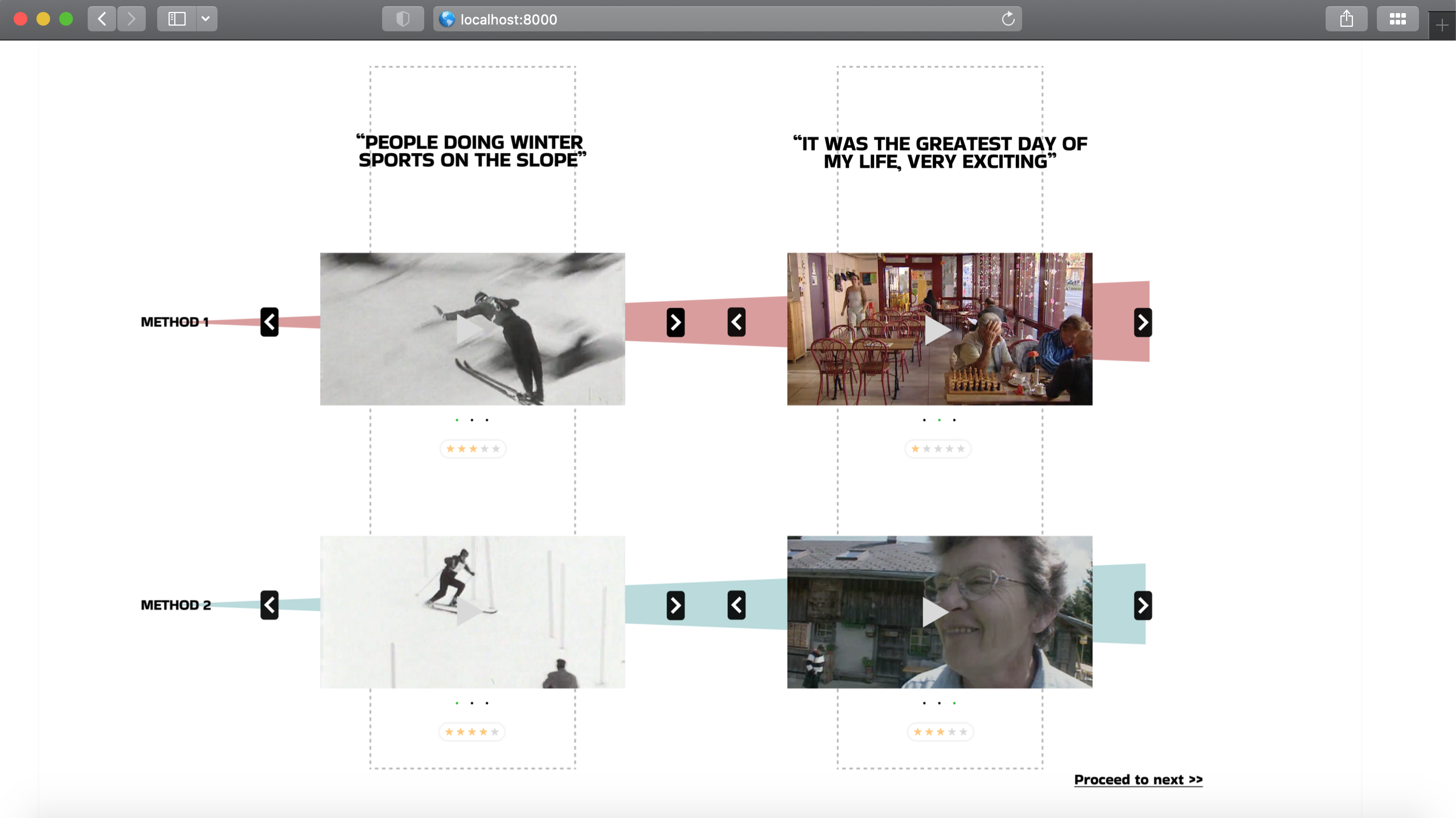} 
    \caption{An example view of the evaluation interface for stage I. The pre-selected visual and quote-focused queries are presented at the top; method 1 represents the customised MFT, and method 2 represents the classifier MFT. Each query results in three videos (as a slide show), and participants are asked to make all clips using the star scale shown under each video.}
\label{6}
\end{figure}

\subsubsection{Stage I}
In the first stage, participants are asked to compare the result of the two methods proposed in the work - customised MFT and classifier MFT. Each screen will show two queries and the top three query results using all three methods (see Fig\ref{6} for an example). The two queries for each page are pre-selected from the 20\% test split, ensuring one query is from captions (focuses on the plain visual) and one from transcripts (focuses on the speech and quote). The participants are asked to compare the relevance of each clip to the query text for a total of ten pages by scoring on a 5-Likert scale. The average score of each method's performance on visual and quote-related queries is calculated and reported for comparison.

\subsubsection{Stage II}
In the second stage, participants are asked to compare the overall text-to-video retrieval with the traditional keyword- and metadata-based retrieval. Participants are advised to recall their past experiences with traditional video retrieval interfaces or explore the typical interface\footnote{\url{https://media.nationalarchives.gov.uk/index.php/category/video/}} provided as a reference. Then, the participants are asked to vote between the overall experiences of the text-to-video retrieval and the traditional search interface based on the acute-eval metric \cite{li2019acute}, which emphasises the four key aspects when evaluating the performance and improvements of novel methods and systems: Engagingness, Interestingness, Humanness, and Informativeness (relevance). For comparison, the sum of votes for text-to-video retrieval and traditional interfaces are reported in the four key aspects.
\section{Results and discussions}
\subsection{Workflow Components}
\subsubsection{Customised MSR-VTT}
\begin{figure}[h]
    \centering 
    \includegraphics[width=.7\columnwidth]{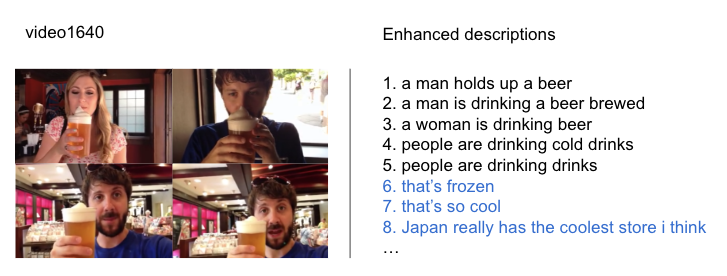} 
    \caption{An example look of the customised MSR-VTT dataset, with random entries in the original descriptions replaced by ASR results when available (in blue).}
\label{fig:3}
\end{figure}

The small model of Whisper obtained a total of 7,302 transcripts out of the 10,000 video clips. The failed cases include situations where the video has no speech or with rare languages. The customised MSR-VTT is then contracted by using the obtained transcripts to replace one or more of the original descriptions paired with a video clip, depending on the availability (see Fig.\ref{fig:3} for an example of one entry in the customised MSR-VTT dataset).

\subsubsection{Classifier}
The classifier used for the classifier-enhanced method is trained on the sequential model in Keras with an 80-20 split for 4000 training and testing to the whole dataset. The sequential model classifier, after training, reached an average accuracy of 97\%, and an average recall of 97\% on the 800-entry test set. 

\subsection{Comparison with state-of-the-art} 
Table \ref{table:modelresult} reports the result of all methods constructed for the evaluation on the comparison with the state-of-the-art. The baseline models' performance on the original MSR-VTT 1k-A train-test split is referenced when available from the original paper. Following the official implementations, the two models are also trained according to the original training settings from scratch using the 1k-A split of the original and the customised MSR-VTT dataset. The test sets from the 1k-A split of the original and the customised MSR-VTT dataset are used to conduct the final evaluation. Overall, the proposed Classifier MFT method achieved comparable performance with the baseline state-of-the-art model, with an R@5 of 54.2 compared to 57.1. On the customised MSR-VTT test set, where the query situation is a bit more complex, Classifier MFT outperforms all other methods with an R@5 of 77.5.

\begin{table}[h]
\begin{tabular}{cccc}
\hline
Method & Training Dataset & Original MSR-VTT & Customised MSR-VTT \\
                 &                    & R@5↑          & R@5↑          \\ \hline
MMT              & Original MSR-VTT   & 54.0          & 12.9          \\
MFT              & Original MSR-VTT   & \textbf{57.1} & 11.2          \\ \hline
Customised MMT   & Customised MSR-VTT & 49.5          & 19.0          \\
Customised MFT    & Customised MSR-VTT & 47.0          & \textbf{22.1} \\ \hline
Classifier MMT   & Original MSR-VTT   & 52.7          & 76.2          \\
Classifier MFT   & Original MSR-VTT                  & \textbf{54.2}    & \textbf{77.5}      \\ \hline
\end{tabular}
\caption{Results of the baseline, customised, and classifier-enhanced method on the original and customised MSR-VTT test sets.}
\label{table:modelresult}
\end{table}

Several observations can be made based on the experiment results. First, all four baseline and customised methods suffer a performance drop when dealing with quote-related queries specific to speech information. This expected behaviour could be caused by the fact that most of the speech information is not matched with the visual perspective of the video clips in the given dataset. Second, models trained on the customised dataset, which contains descriptions that are transcriptions of the video, perform slightly worse when tested with the original test set, but better when dealing with a hybrid of descriptive and quote-related queries. Adding the speech-related descriptions in the customised dataset can provide more information during the training and slightly improve the performance when dealing with queries targeted more on the audio perspective. However, the extra information can also be regarded as noise, messing up the joint-embedding space and undermining the overall performance. Third, all classifier-enhanced methods perform well in both test sets. However, it is noticeable that the performance when dealing with the original test set drops slightly compared to the baseline methods. This can be caused by the fact that the performance is heavily determined by the classifier's performance, in which case it will not be 100\% accurate. 

These results verified that both the customised dataset and the classifier could improve performance to a certain extent when handling more complex query situations. The classifier-enhanced methods are the most promising ones, not only because they reach the highest performance score, but also due to the potential for scalability. The classifier can easily expand to handle more query types and send them to suitable retrieval methods. Overall, this research serves as a valuable foundation to improve the state-of-the-art text-to-video retrieval model's performance in real-world settings and explore innovative and engaging user experiences for accessing large AV archives.

\subsection{Human evaluation}
\subsubsection{RTS dataset}
\begin{figure}
    \centering 
    \includegraphics[width=1\columnwidth]{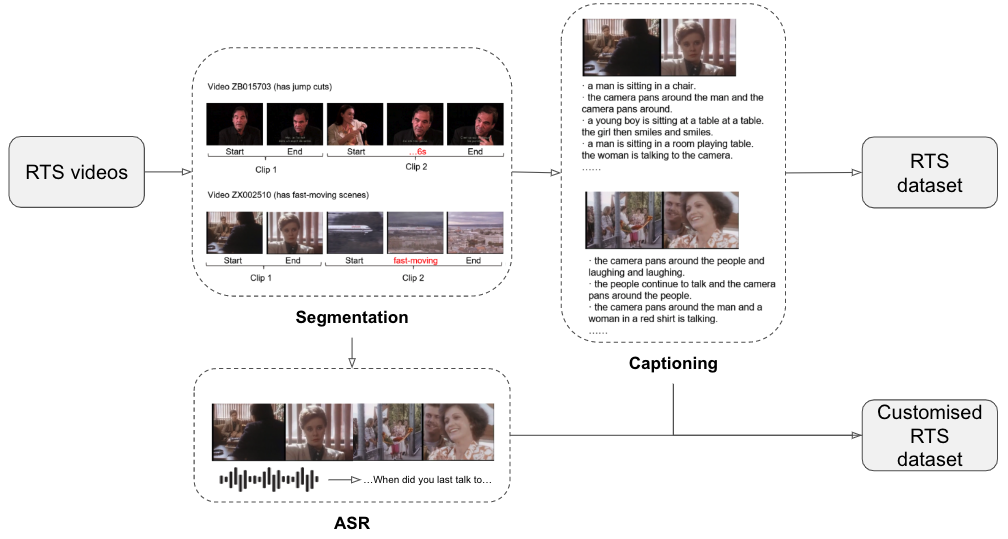} 
    \caption{An overview of the construction of RTS datasets.}
\label{4}
\end{figure}

Fig.\ref{4} provides an overview of the construction of RTS datasets, with example results of each data processing step. Video segmentation
produced a total of 7,348 short clips from the 100 chosen RTS videos. The captioning step produced around 96,000 captions, whereas the ASR step produced around 30,000 transcripts for all the video clips. Table \ref{table:3} provides an overview of the comparison between the RTS datasets and the MSR-VTT.

\begin{table}[h]
\begin{tabular}{cccc}
\hline
                 & RTS dataset & Customised RTS dataset & MSR-VTT \\ \hline
\#Video       & 7,348           & 7,348                  & 10,000                         \\
Avg. duration       & 26.01           & 26.01                  & 9.28                         \\
\#captions \/ video       & 13.1           & 17.1                  & 20                         \\\hline
\end{tabular}
\caption{Comparison between the RTS dataset and the MSR-VTT.}
\label{table:3}
\end{table}

\subsubsection{Evaluation result}
\begin{figure}[h]
    \centering 
    \includegraphics[width=.7\columnwidth]{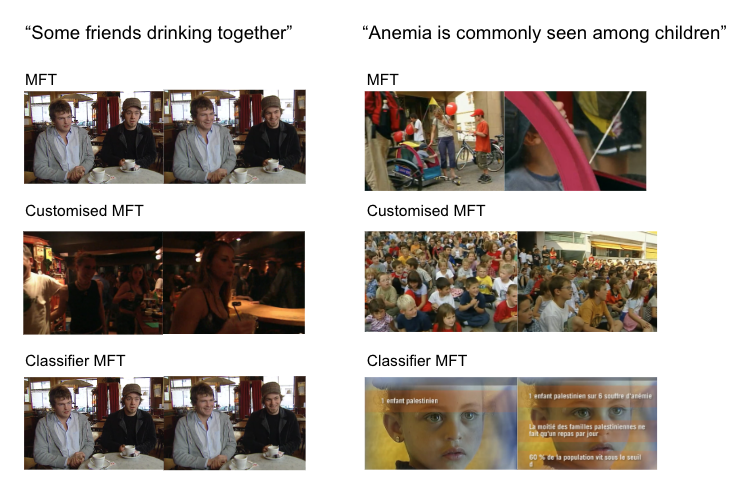} 
    \caption{Examples of the retrieval results using each method and query from the original test set  and is speech-related.}
\label{3}
\end{figure}

Table \ref{table:modelresult} reports the Stage I evaluation results. The three methods performed more or less the same on visual queries, but the Customised MFT performed slightly better. This might be because the customised MFT is trained with the Customised RTS dataset, where descriptions are combined with the ASR transcripts, providing more training. The Classifier MFT is again the best performed on the quote-related query's side. The Stage II evaluation results are reported in Table \ref{table:4}. Participants voted on the four key aspects for evaluating novel interfaces. The text-to-video method wins far ahead in the comparison on three out of the four aspects, namely engagingness, interestingness, and humanness. Regarding informativeness (or relevance), even though the text-to-video method is still deemed better, the vote is very close (6 to 5).  Fig.\ref{3} provides examples of the retrieval results using each method and queries from the original test set (left) and are speech-related (right). The classifier-enhanced method provides better results when the query is speech-related.

The human evaluation provides an alternative perspective to the standard metrics for understanding the performance of text-to-video retrieval models and verifies the classifier-enhanced method works the best. On top of that, applying text-to-video retrieval methods to the real-world data of RTS, this evaluation also testifies to the ability of such models to support a satisfactory experience to explore audiovisual archives.

\begin{table}[h]
\begin{tabular}{cccc}
\hline
                 & Visual Queries & Quote-related Queries \\ \hline
MFT                         & 3.72           & 1.55     \\
Customised MFT              & \textbf{3.90}           & 2.09     \\
Classifier MFT              & 3.64           & \textbf{4.64}     \\ \hline
\end{tabular}
\caption{The average score of each method's performance on visual queries and quote-related queries on stage I human evaluation.}
\label{table:2}
\end{table}

\begin{table}[h]
\begin{tabular}{ccccc}
\hline
                 & Engagingness & Interestingness & Humanness & Informativeness (relevance) \\ \hline
Text-to-video    &  9             & 11            & 10      & 6    \\
Traditional     &  2            & 0            & 1      & 5   \\\hline
\end{tabular}
\caption{The sum of votes for text-to-video retrieval and traditional interfaces in the four key aspects on stage II human evaluation.}
\label{table:4}
\end{table}

\section{Future work}
This work has presented and verified the possible directions of improving existing text-to-video retrieval models to cater to complex real-world query scenarios. However, Based on the current result and scope of the project, the potential key areas to work on in the future are presented as inspirations for similar studies: 

\textbf{Situated datasets.}
Multiple works on text-to-video retrieval methods have emphasised the importance of having a more situated and better quality dataset in improving the retrieval performance \cite{chen2022msr, shvetsova2022everything, fang2021clip2video}. In this specific work, only one additional quote-related query text is considered. However, narrative text, even as simple as a diary, has a much more diverse type of sentence describing many different aspects and levels of semantics within AV content. It would be beneficial to dig deeper in that direction and find a better strategy to create an appropriate customised annotation for video clips to reflect that. For instance, Vlogs, with the speech information being diverse enough to include many aspects of the given video, could be a more suitable source of descriptions for creating the text-video pairs to cover a more diverse scenario in the query \cite{fouhey2018lifestyle}. However, if the more complex annotation will be regarded as noise and hinder the performance is yet to be tested. New training strategies and architectures to handle the weakly-paired text are also required.

\textbf{Intent detection.}
The other focus for planned improvements is on the system end. By refining and improving the classifier to handle more query scenarios and to be able to generate and dispatch queries based on the sentence type and intent to more suitable retrieval models (e.g. quote query to text similarity search, time and place query to metadata boolean search). Going beyond assigning queries to different retrieval models based on sentence type, the intent behind the query can also be inferred. Understanding the intent behind the query text might play a more critical role in providing more relevant results. For example, a description of scenery might be a cue for expressing a certain emotion rather than the scenery itself, and a compliment might be a sarcastic one meant to express hatred towards something. Finding a systematic way to understand the user intent in using different query text would greatly improve the overall experience of the system.

\textbf{Efficiency.}
The system will also implement commonly-adopted post-processing compressing methods for the high-dimensional vectors such as Locality-sensitive hashing \cite{datar2004locality}, Product quantization \cite{jegou2010product}, and Optimized product quantization \cite{ge2013optimized} to deal with the low-efficiency issue brought by the brutal force similarity search, including the text-to-video model used in this speculation. However, such a problem faces inevitable decay when compressing. To merge the compressing to the learning stage and retain more information, one direction to explore is the possibility of developing a quantized representation learning method for text-video retrieval, adapting the work from \cite{wang2022hybrid}.

\section{Conclusion}
This work examined the rising text-to-video retrieval method focusing on real-world applications. We experimented, compared and validated two potential ways to flexibly handle the additional type of query, such as quotes, on top of the basic plain and visually focused ones. While validating the customised model approach can be improved using a more situated dataset, this work favours a more practical method by introducing a classifier as a controller to direct different queries to different retrieval models. This work lays a solid foundation for applying such methods to AV archives for building innovative and explorative interfaces in the future.

\newpage
\bibliographystyle{ACM-Reference-Format}
\bibliography{jocch_2022}

\end{document}